\documentclass[mathleft,
]{an}
\usepackage{graphicx}
\usepackage{times}
\usepackage{natbib}
\bibpunct{(}{)}{;}{a}{}{,}

\usepackage{amsmath}
\usepackage{amssymb}
\newcommand{\bmath}[1]{{\boldsymbol{#1}}}
\newcommand{\rmn}[1]{{\mathrm{#1}}}

\begin{document}

\Yearpublication{2005}%
\Yearsubmission{2005}%
\Month{11}%
\Volume{999}%
\Issue{88}%

\title{On the feasibility of employing solar-like oscillators as
  detectors for the stochastic background of gravitational waves}

\author{D.M. Siegel\inst{1}\fnmsep\thanks{Corresponding author:
  \email{daniel.siegel@aei.mpg.de}\newline}
\and  M. Roth\inst{2}
}
\titlerunning{Solar-like oscillators as detectors for gravitational waves}
\authorrunning{D.M. Siegel \& M. Roth}
\institute{
Max-Planck-Institut f\"ur Gravitationsphysik,
Albert-Einstein-Institut, Am M\"uhlenberg 1, 14476 Potsdam-Golm, Germany
\and 
Kiepenheuer-Institut f\"ur Sonnenphysik, Sch\"oneckstr. 6, 79104
Freiburg, Germany}

\received{2012 Oct 17}
\accepted{2012 Oct 22}
\publonline{2012 Dec 3}


\abstract{We present a hydrodynamic model that describes excitation of
  linear stellar oscillations by a stochastic background of
  gravitational waves (SBGW) of astrophysical and cosmological
  origin. We find that this excitation mechanism is capable of
  generating solar g-mode amplitudes close to or
  comparable with values expected from excitation by turbulent
  convection, which is considered to be the main driving force for
  solar-like oscillations. A method is presented that places direct
  upper bounds on the SBGW in a frequency range, in which the SBGW is
  expected to contain rich astrophysical information. Employing estimates
  for solar g-mode amplitudes, the proposed method is demonstrated
  to have the potential to compete with sensitivities reached by
  gravitational wave experiments in other frequency ranges.}

\maketitle

\section{Introduction}
One of the prime discoveries in cosmology has been the detection of
the cosmic microwave background (CMB) by \cite{Penzias1965}, which captures the state of the
universe roughly 300\,000 years after the Big
Bang, when matter and radiation decoupled from each other and the
universe became transparent to electromagnetic radiation. It is
expected that the universe is also permeated by a stochastic
background of gravitational radiation of astrophysical and
cosmological origin
\citep{Maggiore2000,Sathyaprakash2009}. Proposed cosmological
processes to generate the cosmological component include standard
inflationary models, pre-Big-Bang models and cosmic strings
\citep{Maggiore2000}. The astrophysical component is thought to arise
from the incoherent superposition of a large number of astrophysical
sources that have an accelerated mass distribution with a quadrupole
moment, like compact binary star systems, core collapse supernovae and
rotating neutron stars (\citealt*{Schneider2010}; \citealt{Regimbau2011}). Since the universe has been transparent with
respect to gravitational waves since the time they were first produced by
cosmological processes in the very early universe,
i.e., before $10^{-24}$\,s after the Big Bang, they are ideal carriers of information
on the cosmological and astrophysical processes that generated them
and thus on the state of the universe at these times. This background
encodes information that is not accessible to conventional
observations based on electromagnetic waves and its amplitude has to be
constrained and measured in a variety of
frequency bands, in order to disentangle the signatures of the various contributions.

In the present paper, we show that asteroseismology can place upper
bounds on the amplitude of a stochastic background of gravitational waves (SBGW) in the mHz and $\mu$Hz frequency
range, where the SBGW is likely to be dominated by the astrophysical
component; in this frequency domain, the astrophysical component contains rich astrophysical information, e.g.,
on the physics of compact objects and on star formation history
(\citealt*{Hils1990}; \citealt{Schneider2010}; \citealt{Regimbau2011}). However, it has proven to be difficult to
probe the astrophysical background in the mHz range and it has thus remained essentially unconstrained until
today. Many experiments are currently being proposed to
explore the gravitational wave sky at mHz and $\mu$Hz frequencies, such as the New Gravitational wave
Observatory NGO (a.k.a. eLISA; \citealt{NGO2011}), the DECi-hertz Interferometer
Gravitational wave Observatory DECIGO \citep{Kawamura2011}, or the
torsion-bar antenna TOBA \citep{Ishidoshiro2011}.

The proposed method to constrain the SBGW at mHz and $\mu$Hz
frequencies is based on theoretical work showing that solar-like
oscillations can be excited by gravitational waves \citep{Siegel2011,Siegel2010}. We elaborate on this
topic in Section \ref{sec:excitation}. In particular, we point out that in
the case of the Sun, g modes are more sensitive to an SBGW than
p modes and that solar g-mode
amplitudes in the case of excitation by an SBGW can reach values
comparable with values expected from excitation by turbulent
convection, which is considered to be the main driving force for
oscillations in solar-like stars. Section
\ref{sec:inverse_problem} is devoted to studying the inverse problem of
constraining an SBGW, given asteroseismic data on the surface
velocities of stellar oscillations. In Section \ref{sec:conclusions},
the potential for an asteroseismic experiment to constrain the SBGW in the mHz and $\mu$Hz frequency regime is discussed and
conclusions are presented. Space missions like
CoRoT \citep{Baglin2006} and \textit{Kepler}
\citep{ChristensenDalsgaard2009}, which record asteroseismic data for
a large number of solar-like stars over a wide range of stellar
parameters, but also ground-based telescope networks like SONG
\citep{Grundahl2009} and LCOGT \citep{Shporer2011},
offer a unique possibility to select stars with optimized sets of global stellar
parameters, and to employ these stars as a large array of low-frequency
antennas for the SBGW using the methods presented here.

\section{Excitation of linear stellar oscillations by gravitational waves}
\label{sec:excitation}

\subsection{General hydrodynamic model}
In this Section, we discuss the theoretical model that describes
excitation of solar-like oscillations by an SBGW (cf. \citealt{Siegel2011,Siegel2010} for more details on
the theoretical framework). The physical mechanism underlying this excitation
is the fact that gravitational waves manifest themselves in
oscillating tidal forces; they periodically stretch and compress the
spatial dimensions orthogonal to the direction of propagation in a
quadrupolar pattern, thus imposing stresses on the matter they pass
through. Therefore, only quadrupolar ($l=2$) stellar
oscillations can be excited.

In order to formulate these considerations in a more mathematical
framework, we start from the equations of energy and momentum
conservation in general relativity\footnote{Greek indices take spacetime values 0, 1, 2, 3, whereas Latin indices take spatial values 1, 2, 3 only. Repeated indices are summed over.},
\begin{equation}
	\nabla_\nu T^{\mu\nu}=0, \label{eq:EMC}
\end{equation}
where $T^{\mu\nu}$ denotes the energy--momentum tensor of an ideal
fluid and $\nabla_\mu$ the covariant derivative. Using Fermi normal coordinates, applying the linearized
theory assumption, assuming Newtonian internal motions, and applying
the long wavelength approximation, we obtain from the spatial
components of Eq.\,(\ref{eq:EMC}):
\begin{equation}
	\frac{\partial\rho\bmath{v}}{\partial t}+\bmath{\nabla\cdot}(\rho\bmath{v}\otimes\bmath{v})=-\nabla p+\bmath{f}_{\rm{GW}}, \label{eq:Euler}
\end{equation}
where $t$, $\rho$, $p$ and $\bmath{v}$ denote, respectively, time, density,
pressure, and the internal velocity field of the star. This equation has an additional driving term with respect to the usual
Euler equation in hydrodynamics, $\bmath{f}_{\rm{GW}}(\bmath{x},t)$, with components
\begin{equation}
	f_{\rm{GW}}^i=\frac{1}{2}\rho \ddot{h}^i_{\phantom{i}j} x^j.
\end{equation}
Here, $\bmath{x}$ denotes the position in physical space and
$\ddot{h}_{ij}$ is the second temporal derivative of the gravitational
wave field $h_{ij}(\bmath{x},t)$. Perturbing Eq.\,(\ref{eq:Euler})
around the equilibrium state of the star in the usual way, retaining
only terms up to linear order in the perturbations, decomposing the velocity
field into a component $\bmath{v}_\rmn{osc}$ describing the velocity field due to
oscillations and a component summarizing all other motions, together
with the assumption of incompressible turbulence, we arrive at the
equation of motion,
\begin{equation}
	\rho\left(\frac{\partial^2}{\partial t^2}-\mathcal{L}\right)\bmath{v}_{\rm{osc}}+\mathcal{D}(\bmath{v}_{\rm{osc}})=\frac{\partial}{\partial t}(\bmath{f}_{\rm{Rey}}+\bmath{f}_{\rm{Entr}}+\bmath{f}_{\rm{GW}}). \label{eq:EOM}
\end{equation}
The wave operator $\mathcal{L}$ is a linear differential
operator, whose eigenvalues and eigenfunctions define the oscillation
frequencies and eigenfunctions of the star
\citep{Siegel2011,Unno1989}. 
Furthermore, the linear damping operator $\mathcal{D}$ captures all the damping terms \citep{Siegel2011,Samadi2001}. On the
right-hand side of Eq.\,(\ref{eq:EOM}), we identify the usual Reynold
and Entropy source terms \citep{Siegel2011,Samadi2001}, which describe
excitation of stellar oscillations by turbulent convection, and the
aforementioned driving term due to gravitational waves, which is an
immediate result from the general-relativistic framework and which is
absent in fully Newtonian fluid dynamics. In other words,
Eq.\,(\ref{eq:EOM}) can be seen as a first-order generalization of the
usual Newtonian equation of motion for linear stellar oscillations (cf. \citealt{Samadi2001})
to general relativity.

For the time being we are solely interested in excitation by
gravitational waves, and therefore we ignore the Reynolds and Entropy
source terms. In this case, Eq.\,(\ref{eq:EOM}) can be solved
analytically (even when considering an arbitrary gravitational wave field), and in the case
of an SBGW, we find the following expression for the mean-square
amplitudes of the oscillations:
\begin{equation}
	\big\langle|A_N|^2\big\rangle=\frac{\pi^2}{25}\frac{\chi_N^2}{\eta_N\omega_N I_N^2}H_0^2\Omega_{\rmn{GW}}(\omega_{N}). \label{eq:sq_ampl}
\end{equation}
Here, we made use of the fact that a stellar oscillation mode can be
written as
\begin{equation}
  \bmath{\xi}_N(\bmath{x},t)=A_N(t)\bmath{\xi}(\bmath{x})\rmn{e}^{-\rmn{i}\omega_N t},
\end{equation}
where $A_N$ is a time-dependent, complex amplitude, $N=(n,l,m)$ is an
abridged index for the radial order $n$, harmonic degree $l$, and azimuthal
order $m$ of the mode,
\begin{equation}
	\bmath{\xi}_N(r,\Theta,\phi)=\left[\xi_{r,nl}(r)\bmath{e}_r+\xi_{h,nl}(r)r\nabla\right]Y_{lm}(\Theta,\phi)
\end{equation}
denotes the spatial eigenfunction of the mode, and $\omega_N$ its
frequency. It is remarkable that the right-hand side of
Eq.\,(\ref{eq:sq_ampl}) is essentially given by the product of two
factors, the first of which only depends on stellar properties,
including the mode frequency $\omega_N$, the damping rate $\eta_N$ of the mode, the mode inertia $I_N$,
and a quantity $\chi_N$ that measures the susceptibility of the mode
with respect to excitation by gravitational waves. The second factor
solely depends on the properties of the SBGW, which is entirely
characterized by the normalized, dimensionless function
\citep{Maggiore2000,Allen1999}
\begin{equation}
	\Omega_{\rmn{GW}}(\nu)=\frac{1}{\rho_c}\frac{\rmn{d}\rho_{\rmn{GW}}}{\rmn{d}\ln\nu}, \label{eq:Om}
\end{equation}
which measures the energy density of gravitational waves per unit
logarithmic interval of frequency in units of the present critical
energy density, $\rho_c$, that is needed to close the universe,
\begin{equation}
	\rho_\rmn{c}=\frac{3c^2H_0^2}{8\pi G}.
\end{equation}
Here, $H_0=70$\,km\,s$^{-1}$\,Mpc$^{-1}$ denotes the present Hubble
expansion rate \citep{Komatsu2011}, $c$ the speed
of light, and $G$ the gravitational constant.

\subsection{Numerical results for solar g modes}
\label{sec:num_results}

\begin{figure}
\centering
\includegraphics[width=83mm]{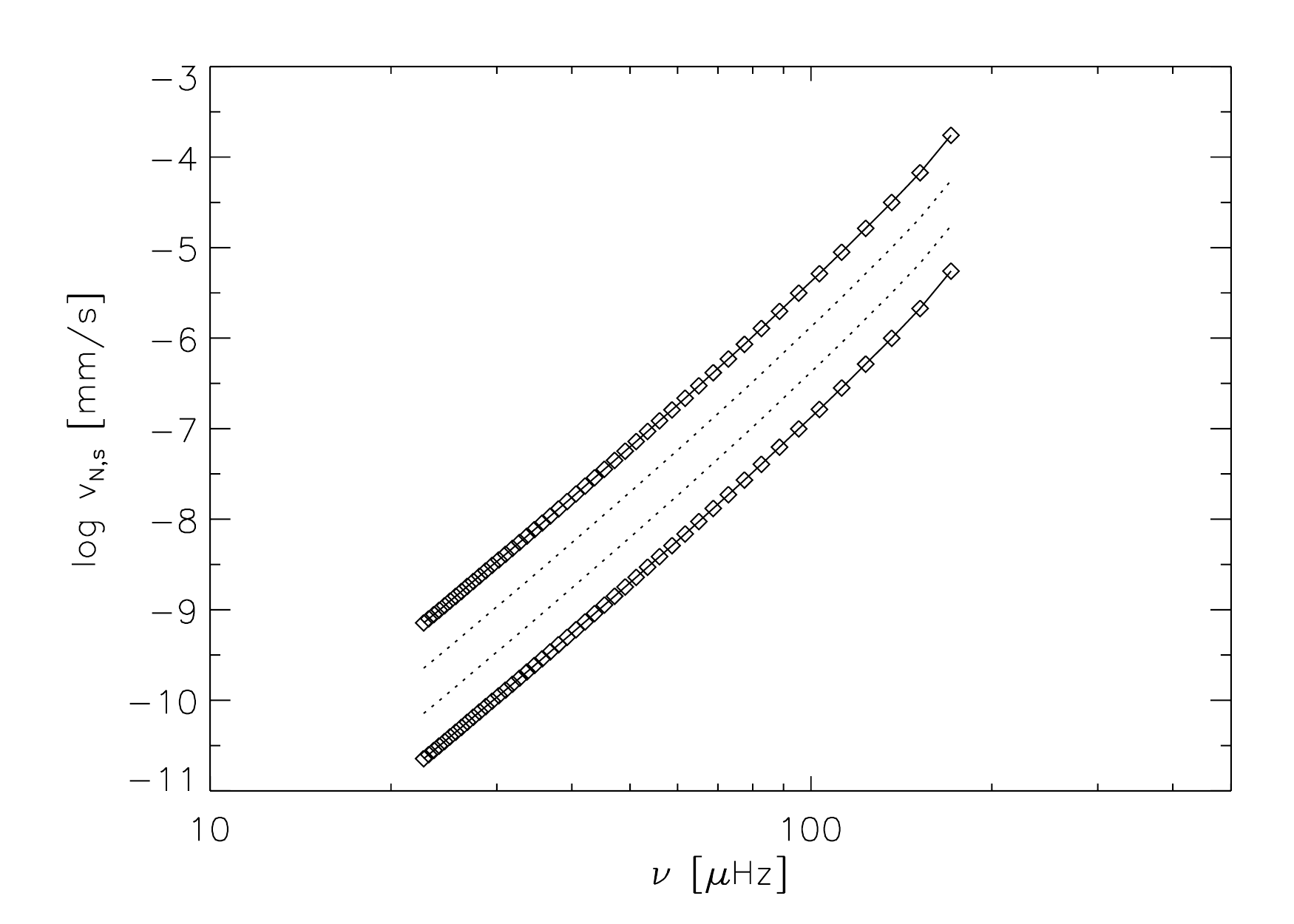}
\caption{Numerical results for intrinsic rms surface velocities of
  $l=2$ solar g modes assuming a constant spectral energy
  density of $\Omega_\rmn{GW} = 1 \times 10^{−5}$ (top curve) and $\Omega_\rmn{GW} = 1 \times
10^{−8}$ (bottom curve) of the SBGW. The dotted curves indicate the cases $\Omega_\rmn{GW} = 1 \times
10^{−6}$ and $\Omega_\rmn{GW} = 1 \times
10^{−7}$ (cf. \citealt{Siegel2011}).}
\label{fig:amplitudes}
\end{figure}

In Figure \ref{fig:amplitudes}, we show numerical results for the
intrinsic root-mean square (rms) surface velocities of quadrupolar solar g modes when excited by an SBGW; as an example, we consider here an SBGW produced by cosmic strings \citep{Siegel2011}. The upper and
lower curve correspond to adopting an optimistic scenario
($\Omega_\rmn{GW} = 1 \times 10^{−5}$ in the considered frequency
range) and a rather pessimistic scenario ($\Omega_\rmn{GW} = 1 \times
10^{−8}$ in the considered frequency range),
respectively. Furthermore, dotted lines indicate the results from setting $\Omega_\rmn{GW} = 1 \times
10^{−6}$ and $\Omega_\rmn{GW} = 1 \times 10^{−7}$. The model
parameters that produce these scenarios are entirely consistent
with all observational constraints on SBGWs to date. Consequently,
depending on model parameters we find maximal surface velocities of $10^{-5}-10^{-3}\,\text{mm}\,\text{s}^{-1}$. 

We note that theoretical estimates for intrinsic rms surface
velocities of $l=1,2,3$ solar g modes in the case of
stochastic excitation by turbulent convection differ from each other
by orders of magnitude. These order of magnitude differences are
predominantly due to the choice of the assumed turbulent eddy-time
correlation function. The maximum values for quadrupolar modes lie
in the range $10^{-3}-1\,\text{mm}\,\text{s}^{-1}$
\citep{Belkacem2009,Appourchaux2010}. Consequently, present models for an SBGW are
capable of exciting quadrupolar solar g modes up to maximum amplitudes that lie very close to (or
possibly even within) the presently expected range from excitation by turbulent
convection.

\section{The inverse problem: constraining an SBGW}
\label{sec:inverse_problem}

In this section, we raise the following question: given observational
data for rms surface velocities of quadrupolar modes, and given that
there is no imprint of an SBGW in the oscillation data, what is the
upper bound on an SBGW at these mode frequencies? If both excitation
mechanisms (turbulent convection and the SBGW) produce rms surface
velocities of the same order of magnitude, one can make use of the
fact that an SBGW only excites quadrupolar modes and observational
data for other harmonic degrees (e.g., $l=1,3$) can be employed to
disentangle the contributions.

From Eq.\,(\ref{eq:sq_ampl}), we derive an upper bound on the
SBGW at the oscillation frequencies of the quadrupolar modes employed,
\begin{equation}
	H_0^2\Omega_\text{GW}(\omega_N)<\frac{50}{\pi^2}\frac{I_N^2}{\Psi_N^2(R)}\frac{\eta_N}{\omega_N}\frac{1}{\chi_N^2}v_{N,s}^2,\label{eq:upper_bound}
\end{equation}
where $v_{N,s}$ denote the intrinsic rms surface velocities and
$\Psi_N^2(R)$ is a quantity that essentially measures the amplitude of the
eigenfunctions at the stellar surface (cf. \citealt{Siegel2011} for details on
this quantity). In deriving
Eq.\,(\ref{eq:upper_bound}), visibility effects have not been
accounted for (e.g., limb darkening, geometrical effects, and height
in the atmosphere where the velocity field is observed); this will be
discussed in a forthcoming paper \citep{Siegel2012}. However,
Eq.\,(\ref{eq:upper_bound}) is accurate in terms of an
order-of-magnitude estimate, and this is what we are interested in for the time being.

In order to investigate the significance of such an asteroseismic bound
on the SBGW, we apply this formalism to the aforementioned case of
quadrupolar solar g modes. The damping rates (as in Section \ref{sec:num_results}) are obtained
from non-adiabatic oscillation computations \citep{Belkacem2009},
which can be reliably calculated for asymptotic g modes due
to the fact that radiative damping is the dominant damping mechanism
for g modes in the asymptotic regime.

\begin{figure}
\centering
\includegraphics[width=83mm]{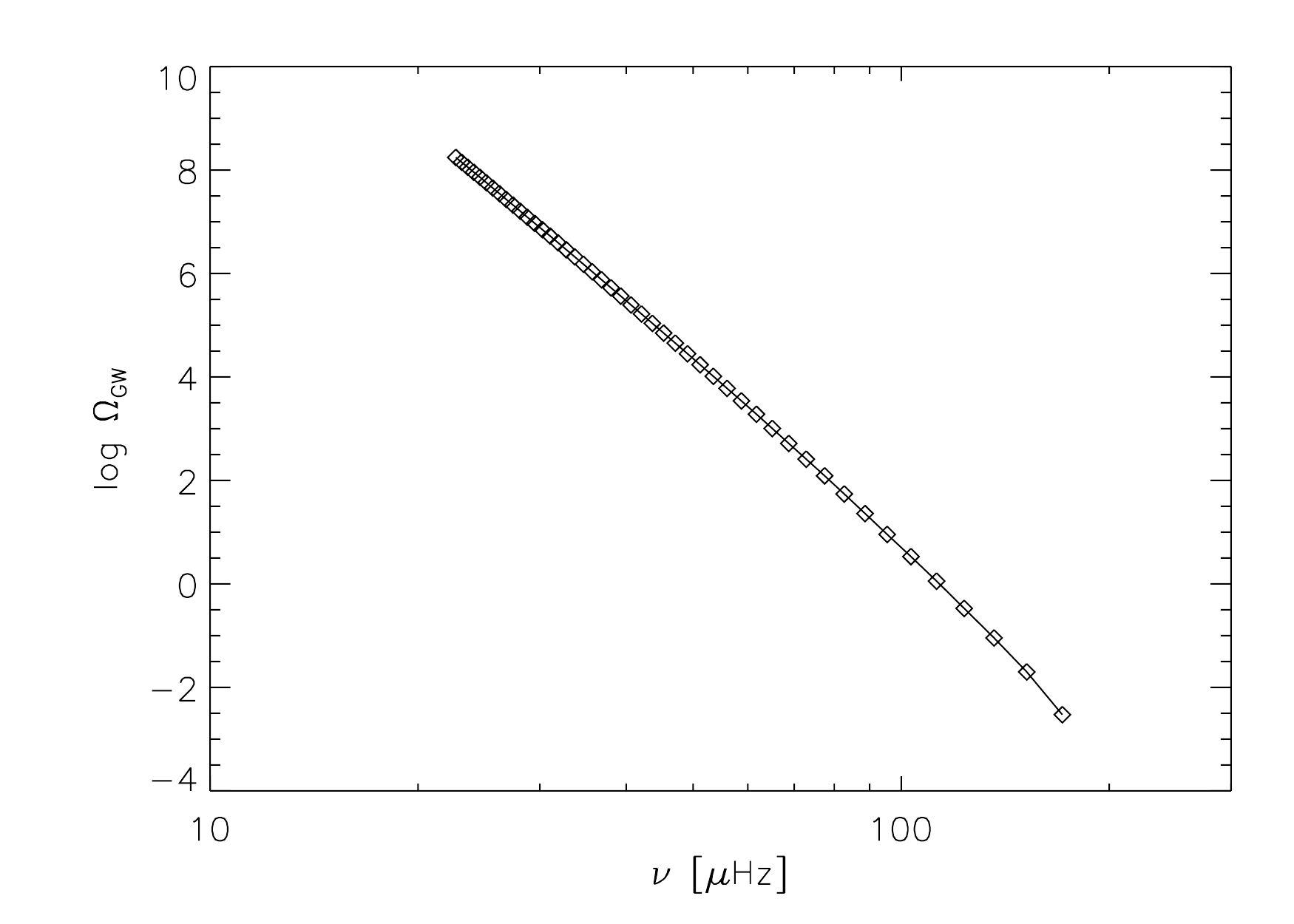}
\caption{Limits deduced from asymptotic, $l=2$ solar g-mode amplitude
  estimates (diamonds indicate the limits from the individual modes) using the
  asteroseismic method reported in this paper (see the text for details).}
\label{fig:upper_bounds}
\end{figure}

Assuming approximately constant rms surface velocities $v_{N,s}
\approx 3\times 10^{-3}$\,mm\,s$^{-1}$ for asymptotic
solar g modes, which corresponds to the calculations of
\citet*{Kumar1996} and \citet{Belkacem2009} in the case of a Gaussian
eddy-time correlation function, we obtain the results depicted in
Figure \ref{fig:upper_bounds}. The power law for the upper
limit on $\Omega_\rmn{GW}(\omega)$ that is evident from Figure
\ref{fig:upper_bounds} is due to the power-law behaviour of the
damping rates for asymptotic solar g modes \citep{Belkacem2009} and
the power laws noticeable in the other quantities in
Eq.\,(\ref{eq:upper_bound}), which also led to the power law for the
rms surface velocities in Figure \ref{fig:amplitudes}. We find that the tightest upper limits
can be deduced from low-order (high-frequency) g modes, with
the most stringent limit being
\begin{equation}
  \Omega_\rmn{GW}\le 3.0\times 10^{-3}\mskip 30mu \text{at}\mskip10mu
  0.17\,\text{mHz}.
  \label{eq:g_mode_ub}
\end{equation}
We note that the upper bound given by Eq.\,(\ref{eq:upper_bound}) is very sensitive to even small changes in
stellar properties and to the observed rms surface velocities
$v_{N,s}$. Assuming, for example, $v_{N,s}\approx 1$\,mm\,s$^{-1}$
results in $\Omega_\rmn{GW}\le 329$ in Eq.\,(\ref{eq:g_mode_ub}).

\section{Conclusions and prospects}
\label{sec:conclusions}

In the present paper, a hydrodynamical model has been presented that describes excitation of
linear stellar oscillations by an SBGW and it has been pointed out
that current models for an SBGW are able to generate solar g-mode
surface amplitudes very close to or comparable with values expected
from excitation by turbulent convection. 

A method has been presented
that is able to directly constrain the energy density of an SBGW at mHz and
$\mu$Hz frequencies, given asteroseismic data on intrinsic rms surface
velocities. This is a so far essentially unexplored frequency regime,
where only indirect bounds on the cosmological component of the SBGW
from CMB and Big Bang nucleosynthesis (BBN) data exist; however, the
SBGW at these frequencies is likely dominated by the astrophysical component. 

Comparing with limits at neighbouring frequencies, the deduced bound based on theoretical estimates for
solar g-mode amplitudes would surpass the limits obtained from Doppler tracking of the
Cassini spacecraft ($10^{-6}-10^{-3}\,\text{Hz}$;
\citealt{Armstrong2003}), the limit from cross-correlation
measurements between the Explorer and Nautilus cryogenic resonant bar
detectors at 907.2\,Hz \citep{Astone1999}, and the bound from the S1 LIGO science run
around 100\,Hz \citep{Abbott2004} by orders of
magnitude. Furthermore, it reaches sensitivities
close to the CMB \citep*{Smith2006}, BBN \citep{Cyburt2005}, LIGO S3,
S4 \citep{Abbott2005,Abbott2007}, and the LIGO S5 (around 100\,Hz;
\citealt{Abbott2009}) bounds. It has thus been demonstrated that the
method presented in this paper is potentially capable of competing with the
sensitivity reached by gravitational wave experiments in other (adjacent)
frequency bands. The upper bounds derived with this method are highly sensitive to local and global stellar properties and missions like \textit{Kepler} and CoRoT offer a
unique possibility to search for targets with an optimized set of
stellar properties, in order to derive tight direct upper limits on an SBGW in the near future.

\acknowledgements
The authors thank K. Belkacem and R. Samadi for valuable
discussions and K. Belkacem for sharing the solar g-mode
damping rates published in \citet{Belkacem2009}. D.M.S. acknowledges
funding from the European Science Foundation for participation in the
conference ``The modern Era of Helio- and Asteroseismology'', which
lead to this proceedings paper. D.M.S. acknowledges funding
from the Max Planck Society and from the Kiepenheuer Institute for
Solar Physics.


\begin{thebibliography}{59}
\bibitem[\protect\citeauthoryear {Abbott et al.}{2004}]{Abbott2004}
  Abbott, B., The LIGO Scientific Collaboration: 2004, Phys. Rev. D 69, 122004
\bibitem[\protect\citeauthoryear {Abbott et al.}{2005}]{Abbott2005}
  Abbott, B., The LIGO Scientific Collaboration: 2005, Phys. Rev. Lett. 95, 221101
\bibitem[\protect\citeauthoryear {Abbott et al.}{2007}]{Abbott2007}
  Abbott, B., The LIGO Scientific Collaboration: 2007, ApJ 659, 918
\bibitem[\protect\citeauthoryear {Abbott et al.}{2009}]{Abbott2009} Abbott, B.P., The LIGO Scientific Collaboration,
  The VIRGO Collaboration: 2009, Nature 460, 990
\bibitem[\protect\citeauthoryear{Allen \& Romano}{1999}]{Allen1999} Allen, B., Romano, J.D.: 1999, Phys. Rev. D 59, 102001
\bibitem[\protect\citeauthoryear{Appourchaux et al.}{2010}]{Appourchaux2010} Appourchaux, T., Belkacem, K., Broomhall, A.-M., et al.: 2010, A\&A Rev. 18, 197
\bibitem[\protect\citeauthoryear{Armstrong et al.}{2003}]{Armstrong2003} Armstrong, J.W., Iess, L., Tortora, P., Bertotti, B.: 2003, ApJ 599, 806
\bibitem[\protect\citeauthoryear{Astone et al.}{1999}]{Astone1999}
  Astone, P., Bassan, M., Bonifazi, P., et al.: 1999, A\&A 351, 811
\bibitem[\protect\citeauthoryear{Baglin et al.}{2006}]{Baglin2006} Baglin, A., Michel, E., Auvergne, M., The COROT
  Team: 2006, in:  K. Fletcher, M. Thompson (eds.),
  \textit{Proceedings of SOHO 18/GONG 2006/HELAS I, Beyond the spherical Sun}, ESA Special Publication 624, p. 34
\bibitem[\protect\citeauthoryear{Belkacem et al.}{2009}]{Belkacem2009}
  Belkacem, K., Samadi, R., Goupil, M.J., Dupret,  M.A., Brun, A.S., Baudin, F.: 2009, A\&A 494, 191
\bibitem[\protect\citeauthoryear{Christensen-Dalsgaard et
    al.}{2009}]{ChristensenDalsgaard2009} Christensen-Dalsgaard, J., Arentoft, T., Brown, T.M., Gilliland, R.L., Kjeldsen, H., Borucki, W.J., Koch, D.: 2009, Comm. in Asteroseismology 158, 328
\bibitem[\protect\citeauthoryear{Cyburt et al.}{2005}]{Cyburt2005} Cyburt, R.H., Fields, B.D., Olive, K.A., Skillman, E.: 2005, Astropart. Phys. 23, 313
\bibitem[\protect\citeauthoryear{Grundahl et al.}{2009}]{Grundahl2009}
  Grundahl, F., Christensen-Dalsgaard, J., Kjeldsen, H., J\o{}rgensen,
  U.G., Arentoft, T., Frandsen, S., Kj\ae{}rgaard, P.: 2009, in: M. Dikpati, T. Arentoft,
  I. Gonz\'alez Hern\'andez, C. Lindsey, F. Hill (eds.), \textit{Solar-Stellar Dynamos as Revealed by Helio- and
    Asteroseismology: GONG 2008/SOHO 21}, ASPC 416, p. 579
\bibitem[\protect\citeauthoryear{Hils, Bender \& Webbink}{Hils et al.}{1990}]{Hils1990} Hils, D., Bender, P.L., Webbink, R.F.: 1990, ApJ 360, 75
\bibitem[\protect\citeauthoryear{Ishidoshiro et
    al.}{2011}]{Ishidoshiro2011} Ishidoshiro, K., Ando, M., Takamori,
  A., et al.: 2011, Phys. Rev. Lett. 106, 161101
\bibitem[\protect\citeauthoryear{Jennrich et al.}{2011}]{NGO2011}
  Jennrich, O., NGO science working team: 2011, ESA/SRE 19, 1
\bibitem[\protect\citeauthoryear{Kawamura et al.}{2011}]{Kawamura2011}
  Kawamura, S., Ando, M., Seto, N., et al.: 2011, Class. Quant. Grav. 28, 094011
\bibitem[\protect\citeauthoryear{Komatsu et al.}{2011}]{Komatsu2011}
  Komatsu, E., Smith, K.M., Dunkleyet, J., et al.: 2011, ApJ Suppl. Ser. 192, 18
\bibitem[\protect\citeauthoryear{Kumar, Quataert \& Bahcall}{Kumar et
    al.}{1996}]{Kumar1996} Kumar, P., Quataert, E.J., Bahcall, J.N.:
  1996, ApJ, 458, L83
\bibitem[\protect\citeauthoryear{Maggiore}{2000}]{Maggiore2000}
  Maggiore, M.: 2000, Phys. Rep. 331, 283
\bibitem[\protect\citeauthoryear{Penzias \&
    Wilson}{1965}]{Penzias1965} Penzias, A.A., Wilson, R.W.: 1965, ApJ
  142, 419
\bibitem[\protect\citeauthoryear{Regimbau}{2011}]{Regimbau2011}
  Regimbau, T.: 2011, Research in Astron. and Astrophys. 11, 369
\bibitem[\protect\citeauthoryear{Samadi \& Goupil}{2001}]{Samadi2001}
  Samadi, R., Goupil, M.-J.: 2001, A\&A 370, 136
\bibitem[\protect\citeauthoryear{Sathyaprakash \& Schutz}{2009}]{Sathyaprakash2009} Sathyaprakash, B.S., Schutz,
  B.F.: 2009, Living Rev. Relativity 12
\bibitem[\protect\citeauthoryear{Schneider, Marassi \&
    Ferrari}{Schneider et al.}{2010}]{Schneider2010} Schneider, R.,
  Marassi, S., Ferrari, V.: 2010, Class. Quant. Grav. 27, 194007
\bibitem[\protect\citeauthoryear{Shporer et al.}{2011}]{Shporer2011}
  Shporer, A., Brown, T., Lister, T., Street, R., Tsapras, Y., Bianco,
  F., Fulton, B., Howell, A.: 2011, in: A. Sozzetti, M.G. Lattanzi,
  A.P. Boss (eds.), IAU Symposium 276, p. 553
\bibitem[\protect\citeauthoryear{Siegel \& Roth}{2010}]{Siegel2010} Siegel, D.M., Roth, M.: 2010, MNRAS 408, 1742
\bibitem[\protect\citeauthoryear{Siegel \& Roth}{2011}]{Siegel2011}
  Siegel, D.M., Roth, M.: 2011, ApJ 729, 137
\bibitem[\protect\citeauthoryear{Siegel \& Roth}{2012}]{Siegel2012}
  Siegel, D.M., Roth, M.: 2012, in prep.
\bibitem[\protect\citeauthoryear{Smith, Pierpaoli \&
    Kamionkowski}{Smith et al.}{2006}]{Smith2006} Smith, T.L., Pierpaoli, E.,
  Kamionkowski, M.: 2006, Phys. Rev. Lett. 97, 021301
\bibitem[\protect\citeauthoryear{Unno et al.}{1989}]{Unno1989} Unno, W., Osaki, Y.,
  Ando, H., Saio, H., Shibahashi, H.: 1989, \textit{Nonradial
    oscillations of stars}, Univ. of Tokyo Press, Tokyo
\end{thebibliography}
\end{document}